\documentclass[10pt]{article}
\usepackage{ametsoc2col}
\newcommand{\myabstract}{The fast-paced development of state-of-the-art limited area models and faster computational resources have made it possible to create simulations at increasing horizontal resolution. This has led to a ubiquitous demand for even higher resolutions from users of various disciplines. This study revisits one of the simulations used in marine ecosystem projects at the Bjerknes Centre. We present a fresh perspective on the assessment of these data, related more specifically to: a) the value added by increased horizontal resolution; and b) a new method for comparing sensitivity studies. The assessment is made using a Bayesian framework for the distribution of mean surface temperature in the Hardanger fjord region in Norway. Population estimates are calculated based on samples from the joint posterior distribution generated using a Monte Carlo procedure. The Bayesian statistical model is applied to output data from the Weather Research and Forecasting (WRF) model at three horizontal resolutions (9, 3 and 1 km) and the ERA Interim Reanalysis. The period considered in this study is from 2007 to 2009, for the months of April, May and June.}
\begin{document}
%
%
\title{\textbf{\large{Horizontal resolution in a nested-domain WRF simulation: a Bayesian analysis approach}}}
%
%
\author{\centerline{\textsc{Michel d. S. Mesquita\footnote{}}}\\
\centerline{\textit{\footnotesize{Uni Climate, Uni Research and Bjerknes Centre for Climate Research, Bergen, Norway}}}\\
\centerline{\textit{\footnotesize{*Corresponding author email: michel.mesquita@uni.no}}}
\and 
\centerline{\textsc{Bj\o rn \AA dlandsvik}}\\
\centerline{\textit{\footnotesize{Institute of Marine Research, Bergen, Norway}}}
\and 
\centerline{\textsc{Cindy Bruy\`{e}re}}\\
\centerline{\textit{\footnotesize{National Center for Atmospheric Research, Boulder,CO, USA}}}
\and 
\centerline{\textsc{Anne D. Sandvik}}\\
\centerline{\textit{\footnotesize{Institute of Marine Research, Bergen, Norway}}}
}
%
\ifthenelse{\boolean{dc}}
{
\twocolumn[
\begin{@twocolumnfalse}
\amstitle

\begin{center}
\begin{minipage}{13.0cm}
\begin{abstract}
	\myabstract
	\newline
	\begin{center}
		\rule{38mm}{0.2mm}
	\end{center}
\end{abstract}
\end{minipage}
\end{center}
\end{@twocolumnfalse}
]
}
{
\amstitle
\begin{abstract}
\myabstract
\end{abstract}
\newpage
}
\section{Introduction}

The need for high-resolution data has become important in several disciplines. These data provide added information, as for example, to the study of complex topography regions such as the Norwegian fjords \citep{heikkilaetal2011,myksvolletal2012}. However, producing such data, using a limited area model, can still be constrained by the computing resources available. For example, in order to make inferences about a model simulation, one needs a large sample to produce robust statistics \citep{lopezetal2006}. Producing large samples at high-resolution can become computationally expensive. This is especially the case when testing different combinations of parameterization schemes or a different model setup.

In this study, we present an alternative approach to analyzing output from limited area models based on Bayesian probability. Bayesian probability theory has been increasingly applied to regional climate modeling experiments in the past few years \citep{buseretal2009,buseretal2010}.  The approach presented here allows one to make use of small samples to make inferences about the statistical population. The use of probability distributions also provides a richer view of the data for comparison against observations. The next session will discuss the data, methods and the Bayesian approach. Section 3 will present the results, which will be followed by the conclusion in Section 4.

\section{Data and Methods}

\begin{figure}[t]
  \noindent\includegraphics[width=19pc,angle=0]{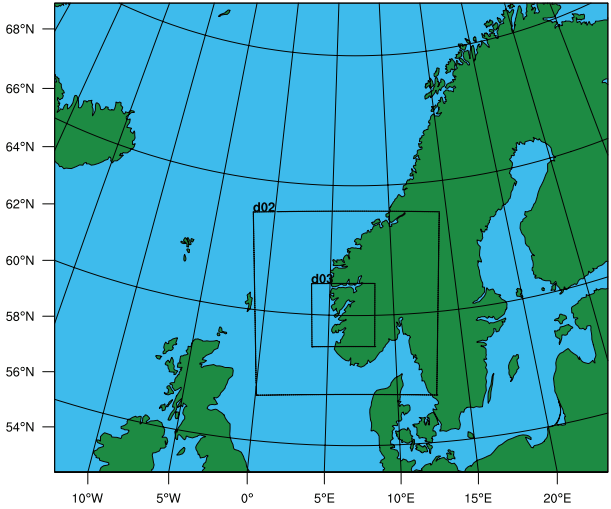}\\
  \caption{WRF model domain setup: parent domain at 9 km (outer domain), nest at 3 km (d02) and nest at 1 km (d03).}\label{f1}
\end{figure}

The experiments were made using the Weather Research and Forecasting (WRF) model version 3.1. Figure \ref{f1} shows the domain configuration, which consisted of a parent domain at 9 km resolution and two nested domains at 3 km and 1 km, respectively (with feedback$=$1, two-way nesting). They were run using 31 vertical levels. The microphysical scheme chosen was the WRF Single-Moment 3-class scheme (mp\_physics$=$3). The cumulus parameterization option was turned off (cu\_physics$=$0). The planetary boundary layer scheme was the Yonsei University scheme (bl\_pbl\_physics$=$1). The longwave radiation scheme used was the RRTM scheme (ra\_lw\_physics$=$1) and the shortwave radiation was the Dudhia scheme (ra\_sw\_physics$=$1).  

ECMWF ERA-interim Re-Analysis was used as the lateral boundary condition data. These data have been obtained from the ECMWF Data Server. The simulation was run from 2007 to 2009. The months of April, May and June of 2008 and 2009 were retained for the analysis. Here, results will be shown with respect to the three-hourly 2 m temperature in the Hardanger fjord region for the month of April. The box selected for the spatial averaging is located between 59.32$^\circ$N, 60.75$^\circ$N and 5.05$^\circ$E, 7.90$^\circ$E. From the timeseries created, we have randomly selected 200 timesteps for calculating the sample mean and variance.

An informative prior was selected based on the Kvams\o y weather station located at 60.358$^\circ$N and 6.275$^\circ$E. These data were obtained from the Norwegian Meteorological Institute data server at $eklima.no$. The Kvams\o y weather station has been operational since November 2003. The average surface temperature for April is 7.48$\pm$1.27$^\circ$C for the years of 2003 to 2011.

\subsection{The Bayesian model}
In this study, the Bayesian model is applied to the 2m temperature in the Hardanger fjord region. It considers the case in which the mean ($\theta$) and variance ($\sigma^2$) are unknown \citep{hoff2009, gelmanetal2004}. For the joint prior distribution $p(\theta,\sigma^2)$ for $\theta$ and $\sigma^2$, the posterior inference will use Bayes' rule, as shown in Equation \ref{eq1}:

\begin{equation}\label{eq1}
p(\theta,\sigma^2\mid y_1,\ldots,y_n)=\frac{p(y_1,\ldots,y_n \mid \theta,\sigma^2)p(\theta,\sigma^2)}{p(y_1,\ldots,y_n)}
\end{equation}

where $y_1,\ldots,y_n$, represent the data. Since the joint distribution for two quantities can be expressed as the product of a conditional probability and a marginal probability, the posterior distribution can likewise be decomposed (Eq. \ref{eq2}):

\begin{equation}\label{eq2}
p(\theta, \sigma^2 \mid y1, \ldots, y_n) = p(\theta \mid \sigma^2, y_1, \ldots, y_n) p(\sigma^2 \mid y_1, \ldots, y_n)
\end{equation}

where the first part of the equation is the conditional probability of $\theta$ on the variance and the data; and the second part is the marginal distribution of $\sigma^2$. The conditional probability part of the equation can be determined as a normal distribution:

\begin{equation}
\{\theta \mid y_1, \ldots, y_n, \sigma^2\} \sim normal(\mu_n, \sigma^2 / \kappa_n)
\end{equation}

Where $\kappa_n=\kappa_0 + n$ represents the degrees of freedom (df) as the sum of the prior df ($\kappa_0$) and that from the data (n). $\mu_n$ is given by: $\mu_n = \frac{(\kappa_0 / \sigma^2)\mu_0 + (n/\sigma^2) \overline{y}}{\kappa_0 / \sigma^2 + n/\sigma^2}=\frac{\kappa_0 \mu_0 + n \overline{y}}{\kappa_n}$, where $\overline{y}$ is the sample mean taken from the WRF simulation. The prior mean is given by $\mu_0$. The calculation of $\sigma^2$ is explained next.

The second part of equation \ref{eq2}, the marginal distribution of $\sigma^2$, can be obtained by integrating over the unknown value of the mean, $\theta$, as follows:

\begin{eqnarray}
p(\sigma^2 \mid y_1, \ldots, y_n) \propto p(\sigma^2) p(y_1, \ldots, y_n \mid \sigma^2) \\
=p(\sigma^2) \int p(y_1, \ldots, y_n \mid \theta, \sigma^2) p(\theta \mid \sigma^2) d\theta
\end{eqnarray}

Solving the integral, and considering the precision ($1/\sigma^2$) such that the distribution is conjugate, gives the following gamma distribution:

\begin{equation}
\{ 1/\sigma^2 \mid y_1, \ldots, y_n  \} \sim gamma(\nu_n/2, \nu_n \sigma_n^2/2) 
\end{equation}

Where $\nu_n = \nu_0 +n$ is the sum of degrees of freedom of the prior ($\nu_0$) and of the data (n). $\sigma_n^2$ is given by $\sigma_n^2 = \frac{1}{\nu_n}[\nu_0 \sigma_0^2 + (n-1) s^2 + \frac{\kappa_0 n}{\kappa_n} (\overline{y} - \mu_0)^2]$, where $\overline{y}$ is the sample mean and $s^2$ is the sample variance, both taken from the WRF simulation. $\sigma_0^2$ is the prior variance.

\subsection{Monte Carlo sampling}

Samples of $\theta$ and $\sigma^2$ can be generated from their joint posterior distribution using the following Monte Carlo procedure \citep{hoff2009}:

\begin{equation*}
\sigma^{2(1)} \sim inv\; gamma(\frac{\nu_n}{2}, \frac{\sigma^2_n \nu_n}{2}),   \quad     \theta^{(1)} \sim normal(\mu_n, \frac{\sigma^{2(1)}}{\kappa_n}) \\
\end{equation*}
$\vdots\\$
\begin{equation*}
\sigma^{2(S)} \sim inv\; gamma(\frac{\nu_n}{2}, \frac{\sigma^2_n \nu_n}{2}),    \quad   \theta^{(S)} \sim normal(\mu_n, \frac{\sigma^{2(S)}}{\kappa_n})
\end{equation*}

where $\sigma^2$ is estimated using an inverse-gamma distribution ($inv\;gamma$). Each $\theta^{(S)}$ is sampled from its conditional distribution given the data and $\sigma^2=\sigma^{2(S)}$. The simulated pairs of $\{(\sigma^{2(1)}, \theta^{(1)}), \ldots, (\sigma^{2(S)}, \theta^{(S)}) \}$ are independent samples of the joint posterior distribution, i.e.: $p(\theta, \sigma^2 \mid y_1, \ldots, y_n)$. The simulated sequence $\{\theta^{(1)}, \ldots, \theta^{(S)}\}$ can be seen as independent samples from the marginal posterior distribution of $p(\theta \mid y_1, \ldots, y_n)$, and so this sequence can be used to make Monte Carlo approximations to functions involving $p(\theta \mid y_1, \ldots, y_n)$. While $\theta^{(1)}, \ldots, \theta^{(S)}$ are each conditional samples, they are also each conditional on different values of $\sigma^2$. Together, they make up marginal samples of $\theta$.

\section{Results}

\begin{figure}[t]
  \noindent\includegraphics[width=19pc,angle=0]{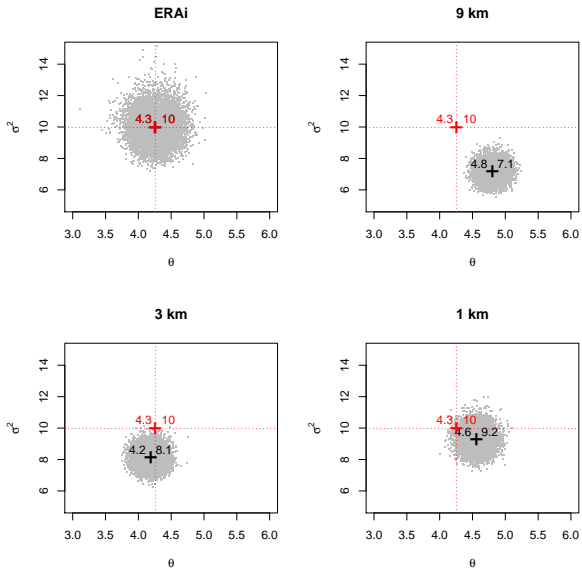}\\
  \caption{Monte Carlo samples from the joint distributions of the population mean ($\theta$) and variance ($\sigma^2$) for ERA Interim (ERAi) and for the different domains. The values in black show the mean value of the population mean (right side) and of the population variance (left side). Accordingly, the mean values of $\theta$ and $\sigma^2$ for the ERA Interim are indicated in red. Temperature given in degrees Celsius.}\label{f2}
\end{figure}

Monte Carlo samples from the joint distributions of the population mean and variance are shown in Figure \ref{f2}. The ERA Interim distribution (ERAi), on the top left, shows larger spread both for the mean and the variance as compared to the three domains. The distribution for the 9 km domain seems to be off and does not match the ERA Interim data. The 3 km nest shows the closest approximation to the mean of the ERA Interim, whereas the 1 km nest approximates the variance more closely. 

\begin{figure}[t]
  \noindent\includegraphics[width=19pc,angle=0]{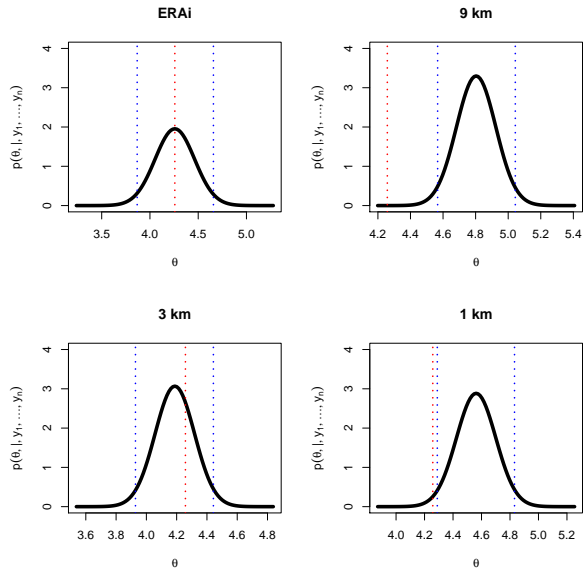}\\
  \caption{Monte Carlo samples from the marginal distribution of $\theta$ for ERA Interim (ERAi) and for the different domains. The blue vertical lines give a 95\% quantile-based posterior bound. In red, the mean value of the ERA Interim posterior marginal distribution. Temperature given in degrees Celsius.}\label{f3}
\end{figure}

Figure \ref{f3} shows the marginal distribution of the mean, based on the Monte Carlo sampling. The red line indicates the mean value of the marginal distribution for the ERA Interim. The posterior bounds of the 9 km parent domain do not contain the mean value of the ERA Interim.  Table \ref{t1} shows that even though there is some overlap between the posterior bounds of ERA Interim and the 9 km domain, this overlap is minimum. The 3 km and 1 km nests show a closer overlap with the ERA Interim data. The 3 km resolution domain is able to approximate the mean more realistically, also confirmed by the posterior bound overlap with ERA Interim (Table \ref{t1}).

\begin{table}[t]
\caption{Posterior distribution summary for the mean ($\theta$) and variance ($\sigma^2$) based on Monte Carlo sampling. The 95\% posterior bound (PB) is also indicated for each variable. Temperature units given in degrees Celsius.}\label{t1}
\begin{center}
\begin{tabular}{cccccrcrc}
\hline\hline
 & $\theta$ & $\theta$ PB & $\sigma^2$ & $\sigma^2$ PB\\
\hline
 ERAi & 4.26 & (3.87, 4.66) & 9.90 & (8.30, 11.93) \\
 d01 & 4.80 & (4.57, 5.04) & 7.08 & (6.26, 8.07) \\
 d02 & 4.19 & (3.93, 4.44) & 8.04 & (7.12, 9.14) \\
 d03 & 4.56 & (4.29, 4.83) & 9.19 & (8.11, 10.46) \\
\hline
\end{tabular}
\end{center}
\end{table}

\begin{figure}[t]
  \noindent\includegraphics[width=19pc,angle=0]{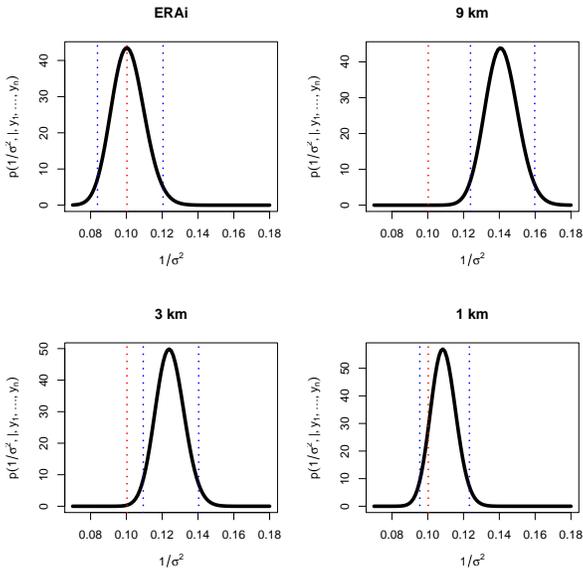}\\
  \caption{The same as Figure \ref{f3}, but for the precision, $1/\sigma^2$.}\label{f4}
\end{figure}

 The marginal distribution of the ERA Interim variance is approximated more closely by the 1 km resolution domain, as shown in Figure \ref{f4}. The mean value of the ERA Interim marginal distribution is within the posterior bounds for that resolution. In contrast, the 9 km and 3 km domains have posterior bounds outside of the ERA Interim mean value. There is, however, a better overlap between the ERA Interim and the 3 km posterior distribution, compared to the 9 km one (Table \ref{t1}).

\section{Conclusion}

This study has used a Bayesian statistical model applied to output data from the Weather Research and Forecasting (WRF) model at three horizontal resolutions (9, 3 and 1 km). Station-based observational data was used to provide an informative prior. We have presented a fresh perspective on the assessment of data from the WRF model, related more specifically to: a) the value added by increased horizontal resolution; and b) a new method for comparing sensitivity studies.

The increased horizontal resolution is able to approximate the mean and the variance of the observations more closely. This approximation is crucial, for example, when one is to use these data to force a regional oceal model \citep{myksvolletal2012}. The Bayesian method introduced here provides a richer probabilistic view of the dataset. It also obviates the use of long simulations for estimating the population mean or variance - thus saving computational resources. In high-resolution experiments such as this, one is constrained by the amount of computational resources used. If one is to use standard statistics, a larger sample is needed to be able to make robust inferences. Hence, through the use of prior information, the Bayesian framework provides an alternative approach to estimating the statistical population, and in this case, for assessing the bias in the model simulation. It is also useful for sensitivity studies where one needs to compare not only resolution, but also the use of different parameterization schemes. This approach can also be applied to other variables by adapting it to their underlying distribution.

\begin{acknowledgment} 
We would like to thank NCAR for making the WRF model publicly available. We also thank ECMWF and the Norwegian Meteorological Institute for the datasets provided. This study has been funded through the Downscaling Synthesis project at the Bjerknes Centre for Climate Research, Bergen, Norway. 
\end{acknowledgment}

\ifthenelse{\boolean{dc}}
{}
{\clearpage}
\bibliographystyle{ametsoc}
\bibliography{references}

\end{document}